\documentclass[10pt]{article}
\usepackage{graphicx}
\usepackage{amsmath}
\usepackage{amssymb}
\usepackage{caption2}
\begin{document}
\begin{center}
{\large {\bf \sc{  Analysis of strong decays of the $Z_c(4600)$ with the QCD sum rules  }}} \\[2mm]
Zhi-Gang  Wang \footnote{E-mail: zgwang@aliyun.com.  }    \\
 Department of Physics, North China Electric Power University, Baoding 071003, P. R. China
\end{center}

\begin{abstract}
In this article, we tentatively assign the $Z_c(4600)$ to be the $[dc]_P[\bar{u}\bar{c}]_A-[dc]_A[\bar{u}\bar{c}]_P$ type vector tetraquark state and study
its two-body strong decays with the QCD sum rules based on solid quark-hadron duality, the predictions for the partial decay widths $\Gamma(Z_c^-\to J/\psi \pi^-)=41.4^{+20.5}_{-14.9}\,\rm{MeV}$, $\Gamma(Z_c^-\to \eta_c\rho^-)=41.6^{+32.7}_{-22.2}\,\rm{MeV}$,
$\Gamma(Z_c^-\to J/\psi a_0^-(980))=10.2^{+11.3}_{-6.7}\,\rm{MeV}$, $\Gamma(Z_c^-\to \chi_{c0}\rho^-)=3.5^{+6.7}_{-3.0}\,\rm{MeV}$,
$\Gamma(Z_c^-\to D^{*0} D^{*-})=39.5^{+29.6}_{-19.3}\,\rm{MeV}$, $\Gamma(Z_c^-\to D^{0} D^{-})=6.6^{+4.6}_{-3.0}\,\rm{MeV}$ and
$\Gamma(Z_c^-\to D^{*0} D^{-})=1.0^{+1.0}_{-0.6}\,\rm{MeV}$ can be confronted to the experimental data in the future to diagnose the nature of the  $Z_c(4600)$.
\end{abstract}

PACS number: 12.39.Mk, 12.38.Lg

Key words: Tetraquark  state, QCD sum rules

\section{Introduction}

Recently, the LHCb collaboration performed an angular analysis of the decays $B^0\to J/\psi K^+\pi^-$ using proton-proton collision data corresponding to an integrated luminosity of $3\rm{fb}^{-1}$  collected with the LHCb detector, studied  the $m(J/\psi \pi^-)$ versus $m(K^+\pi^-)$ plane, and observed two possible
 structures near $m(J/\psi \pi^-)=4200 \,\rm{MeV}$ and $4600\,\rm{MeV}$, respectively \cite{LHCb-Z4600}. The structure near $m(J/\psi \pi^-)=4200 \,\rm{MeV}$ is
 close to the exotic state $Z_c(4200)$ reported previously by the Belle collaboration \cite{Belle-4200}. The structure near $m(J/\psi \pi^-)=4600 \,\rm{MeV}$ is in excellent agreement with our previous prediction of the mass of the $[qc]_P[\bar{q}\bar{c}]_A-[qc]_A[\bar{q}\bar{c}]_P$  type tetraquark state with the spin-parity-charge-conjugation $J^{PC}=1^{--}$, $m_Y=(4.59 \pm 0.08)\,\rm{GeV}$ \cite{WangY4360Y4660-1803}.

 In Ref.\cite{WangY4360Y4660-1803}, we perform detailed and updated analysis of the
  $[sc]_S[\bar{s}\bar{c}]_V+[sc]_V[\bar{s}\bar{c}]_S$ type, $[qc]_S[\bar{q}\bar{c}]_V+[qc]_V[\bar{q}\bar{c}]_S$ type,
$[sc]_P[\bar{s}\bar{c}]_A-[sc]_A[\bar{s}\bar{c}]_P$ type and $[qc]_P[\bar{q}\bar{c}]_A-[qc]_A[\bar{q}\bar{c}]_P$ type vector tetraquark states with the QCD sum rules based on our previous works \cite{Wang-tetra-formula,WangEPJC-1601},
  the predictions  support assigning  the $Y(4660)$  to be the  $[sc]_P[\bar{s}\bar{c}]_A-[sc]_A[\bar{s}\bar{c}]_P$ type vector tetraquark state, assigning the $Y(4360/4320)$ to be the $[qc]_S[\bar{q}\bar{c}]_V+[qc]_V[\bar{q}\bar{c}]_S$   type vector tetraquark state.
 In Ref.\cite{Wang-Y4660-Decay}, we  choose  the $[sc]_P[\bar{s}\bar{c}]_A-[sc]_A[\bar{s}\bar{c}]_P$  type tetraquark current to study the hadronic coupling constants in the strong decays $Y(4660)\to J/\psi f_0(980)$, $ \eta_c \phi(1020)$,    $ \chi_{c0}\phi(1020)$, $ D_s \bar{D}_s$, $ D_s^* \bar{D}^*_s$, $ D_s \bar{D}^*_s$,  $ D_s^* \bar{D}_s$, $  \psi^\prime \pi^+\pi^-$, $J/\psi\phi(1020)$ with the QCD sum rules based on solid quark-hadron duality. The predicted width $\Gamma(Y(4660) )= 74.2^{+29.2}_{-19.2}\,{\rm{MeV}}$ is in excellent agreement with the experimental data $68\pm 11\pm 1 {\mbox{ MeV}}$ from the Belle collaboration \cite{Belle-Y4660-2014}, which also supports assigning the $Y(4660)$ to be the  $[sc]_P[\bar{s}\bar{c}]_A-[sc]_A[\bar{s}\bar{c}]_P$  type tetraquark state with $J^{PC}=1^{--}$.
 In the isospin limit, the tetraquark states with the symbolic quark structures $c\bar{c} u\bar{d}$, $c\bar{c} d\bar{u}$, $c\bar{c} \frac{u\bar{u}-d\bar{d}}{\sqrt{2}}$, $c\bar{c} \frac{u\bar{u}+d\bar{d}}{\sqrt{2}}$ have degenerated masses.

 In this article, we tentatively assign the $Z_c(4600)$ to be the $[dc]_P[\bar{u}\bar{c}]_A-[dc]_A[\bar{u}\bar{c}]_P$ type vector tetraquark state and study
 its two-body  strong decays with the QCD sum rules based on solid quark-hadron duality by taking   into account both the connected and disconnected Feynman diagrams in the operator product expansion \cite{WangZhang-Solid}, which is valuable in understanding the nature of the vector tetraquark states.

The article is arranged as follows:  we  obtain the QCD sum rules for  the hadronic coupling constants
 $G_{Z_cJ/\psi \pi}$,   $ G_{Z_c\eta_c \rho}$, $G_{Z_cJ/\psi a_0}$,  $G_{Z_c\chi_{c0} \rho}$, $G_{Z_c\bar{D}^*D^*}$,  $G_{Z_c\bar{D}D}$ and
$G_{Z_c\bar{D}^*D}$   in section 2;  we   present the numerical results and discussions in section 3; section 4 is reserved for our conclusion.

\section{The QCD sum rules for the hadronic coupling constants}

Now we write down the  three-point correlation functions for    the hadronic coupling constants
 $G_{Z_cJ/\psi \pi}$,   $ G_{Z_c\eta_c \rho}$, $G_{Z_cJ/\psi a_0}$,  $G_{Z_c\chi_{c0} \rho}$, $G_{Z_c\bar{D}^*D^*}$,  $G_{Z_c\bar{D}D}$ and $G_{Z_c\bar{D}^*D}$ in the two-body strong decays $Z_c(4600) \to J/\psi \pi$,   $\eta_c \rho$, $J/\psi a_0$,  $\chi_{c0} \rho$, $D^*\bar{D}^*$,  $D\bar{D}$, $D^*\bar{D}$ and $D\bar{D}^*$, respectively,
\begin{eqnarray}
\Pi^1_{\mu\nu}(p,q)&=&i^2\int d^4xd^4y e^{ipx}e^{iqy}\langle 0|T\Big\{J_{J/\psi,\mu}(x)J_{\pi}(y)J_{\nu}^{\dagger}(0)\Big\}|0\rangle\, , \\
\Pi^2_{\mu\nu}(p,q)&=&i^2\int d^4xd^4y e^{ipx}e^{iqy}\langle 0|T\Big\{J_{\eta_c}(x)J_{\rho,\mu}(y)J_{\nu}^{\dagger}(0)\Big\}|0\rangle\, ,\\
\Pi^3_{\mu\nu}(p,q)&=&i^2\int d^4xd^4y e^{ipx}e^{iqy}\langle 0|T\Big\{J_{J/\psi,\mu}(x)J_{a_0}(y)J_{\nu}^{\dagger}(0)\Big\}|0\rangle\, ,\\
\Pi^4_{\mu\nu}(p,q)&=&i^2\int d^4xd^4y e^{ipx}e^{iqy}\langle 0|T\Big\{J_{\chi_{c0}}(x)J_{\rho,\mu}(y)J_{\nu}^{\dagger}(0)\Big\}|0\rangle\, ,\\
\Pi^5_{\alpha\beta\nu}(p,q)&=&i^2\int d^4xd^4y e^{ipx}e^{iqy}\langle 0|T\Big\{J^{\dagger}_{D^{0*},\alpha}(x)J_{D^{*+},\beta}(y)J_{\nu}^{\dagger}(0)\Big\}|0\rangle\, ,\\
\Pi^6_{\nu}(p,q)&=&i^2\int d^4xd^4y e^{ipx}e^{iqy}\langle 0|T\Big\{J^{\dagger}_{D^0}(x)J_{D^+}(y)J_{\nu}^{\dagger}(0)\Big\}|0\rangle\, ,\\
\Pi^7_{\mu\nu}(p,q)&=&i^2\int d^4xd^4y e^{ipx}e^{iqy}\langle 0|T\Big\{J^{\dagger}_{D^{*0},\mu}(x)J_{D^+}(y)J_{\nu}^{\dagger}(0)\Big\}|0\rangle\, ,
\end{eqnarray}
where the  currents,
\begin{eqnarray}
J_{J/\psi,\mu}(x)&=&\bar{c}(x)\gamma_\mu c(x)\, ,\nonumber\\
J_{\rho,\mu}(y)&=&\bar{d}(y)\gamma_\mu  u(y)\, ,\nonumber\\
J_{\eta_c}(x)&=&\bar{c}(x)i\gamma_5 c(x)\, ,\nonumber\\
J_{\pi}(y)&=&\bar{d}(y)i\gamma_5  u(y)\, ,\nonumber\\
J_{a_0}(y)&=&\bar{d}(y)  u(y)\, ,\nonumber\\
J_{\chi_{c0}}(x)&=&\bar{c}(x) c(x)\, ,\nonumber\\
J_{D^+}(y)&=&\bar{d}(y)i\gamma_5 c(y)\, ,\nonumber\\
J_{D^0}(x)&=&\bar{u}(x)i\gamma_5 c(x)\, ,\nonumber\\
J_{D^{*+},\beta}(y)&=&\bar{d}(y)\gamma_\beta c(y)\, ,\nonumber\\
J_{D^{*0},\alpha}(x)&=&\bar{u}(x)\gamma_\alpha c(x)\, ,\nonumber\\
J_\nu(0)&=&\frac{\varepsilon^{ijk}\varepsilon^{imn}}{\sqrt{2}}\Big[u^{Tj}(0)C c^k(0) \bar{d}^m(0)\gamma_\nu C \bar{c}^{Tn}(0)-u^{Tj}(0)C\gamma_\nu c^k(0)\bar{d}^m(0)C \bar{c}^{Tn}(0) \Big] \, ,
\end{eqnarray}
we have assumed that the dominant components of the $a_0(980)$ are two-quark states \cite{WangScalarNonet}.
We choose  those currents shown in Eq.(8) to interpolate the corresponding mesons according to the standard definitions for the  current-hadron couplings or the decay constants $f_{J/\psi}$,
$f_\rho$, $f_{\eta_c}$, $f_{\pi}$, $f_{a_0}$, $f_{\chi_{c0}}$, $f_{D}$, $f_{D^*}$, $\lambda_Z$,
\begin{eqnarray}
\langle 0|J_{J/\psi,\mu}(0)|J/\psi(p)\rangle&=&f_{J/\psi}m_{J/\psi}\xi^{J/\psi}_\mu\, ,\nonumber\\
\langle 0|J_{\rho,\mu}(0)|\rho(p)\rangle&=&f_{\rho}m_{\rho}\xi^{\rho}_\mu\, ,\nonumber\\
\langle 0|J_{\eta_c}(0)|\eta_c(p)\rangle&=&\frac{f_{\eta_c}m_{\eta_c}^2}{2m_c}\, ,\nonumber\\
\langle 0|J_{\pi}(0)|\pi(p)\rangle&=&\frac{f_{\pi}m_{\pi}^2}{m_u+m_d}\, ,\nonumber\\
\langle 0|J_{a_0}(0)|a_0(p)\rangle&=&f_{a_0}m_{a_0}\, ,\nonumber\\
\langle 0|J_{\chi_{c0}}(0)|\chi_{c0}(p)\rangle&=&f_{\chi_{c0}}m_{\chi_{c0}}\, ,\nonumber\\
\langle 0|J_{D}(0)|D(p)\rangle&=&\frac{f_{D}m_{D}^2}{m_c}\, ,\nonumber\\
\langle 0|J_{D^*,\mu}(0)|D^*(p)\rangle&=&f_{D^*}m_{D^*}\xi^{D^*}_\mu\, ,\nonumber\\
\langle 0|J_{\mu}(0)|Z_c(p)\rangle&=&\lambda_{Z}\xi^{Z}_\mu\, ,
\end{eqnarray}
where the $\xi_\mu$ are the polarization vectors.

At the hadronic   side,  we insert  a complete set of intermediate hadronic states with the same quantum numbers as the current operators  into the three-point
correlation functions  and  isolate the ground state contributions to obtain the  following results \cite{SVZ79-1,SVZ79-2,Reinders85},

\begin{eqnarray}
\Pi^1_{\mu\nu}(p,q)&=& \frac{f_{\pi}m_{\pi}^2}{m_u+m_d}\frac{f_{J/\psi}m_{J/\psi}\,\lambda_Z\,G_{Z_cJ/\psi\pi}\,\varepsilon_{\alpha\beta\rho\sigma}p^\alpha p^{\prime\rho}}{\left(p^{\prime2}-m_Z^2\right)\left(p^2-m_{J/\psi}^2 \right)\left(q^2-m_{\pi}^2\right)}\left(-g_{\mu}{}^{\beta}+\frac{p_{\mu}p^{\beta}}{p^2}\right)\left(-g_\nu{}^\sigma+\frac{p^\prime_{\nu}p^{\prime\sigma}}{p^{\prime2}}\right)+\cdots\nonumber\\
&=&\Pi(p^{\prime2},p^2,q^2)\,\left(-\varepsilon_{\mu\nu\alpha\beta}p^{\alpha}q^{\beta}\right)+\cdots \, ,
\end{eqnarray}

\begin{eqnarray}
\Pi^2_{\mu\nu}(p,q)&=& \frac{f_{\eta_c}m_{\eta_c}^2}{2m_c}\frac{f_{\rho}m_{\rho}\,\lambda_Z\,G_{Z_c\eta_c\rho}\,\varepsilon_{\alpha\beta\rho\sigma}q^\alpha p^{\prime\rho}}{\left(p^{\prime2}-m_Z^2\right)\left(p^2-m_{\eta_c}^2 \right)\left(q^2-m_{\rho}^2\right)}\left(-g_{\mu}{}^{\beta}+\frac{q_{\mu}q^{\beta}}{q^2}\right)\left(-g_\nu{}^\sigma+\frac{p^\prime_{\nu}p^{\prime\sigma}}{p^{\prime2}}\right)+\cdots\nonumber\\
&=&\Pi(p^{\prime2},p^2,q^2)\,\varepsilon_{\mu\nu\alpha\beta}p^{\alpha}q^{\beta}+\cdots \, ,
\end{eqnarray}

\begin{eqnarray}
\Pi^3_{\mu\nu}(p,q)&=& \frac{f_{J/\psi}m_{J/\psi}f_{a_0}m_{a_0}\,\lambda_Z\,G_{Z_cJ/\psi a_0}}{\left(p^{\prime2}-m_Z^2\right)\left(p^2-m_{J/\psi}^2 \right)\left(q^2-m_{a_0}^2\right)}\left(-g_{\mu\alpha}+\frac{p_{\mu}p_{\alpha}}{p^2}\right)\left(-g_\nu{}^\alpha+\frac{p^\prime_{\nu}p^{\prime\alpha}}{p^{\prime2}}\right)
+\cdots\nonumber\\
&=&\Pi(p^{\prime2},p^2,q^2)\,g_{\mu\nu}+\cdots \, ,
\end{eqnarray}

\begin{eqnarray}
\Pi^4_{\mu\nu}(p,q)&=& \frac{f_{\chi_{c0}}m_{\chi_{c0}}f_{\rho}m_{\rho}\,\lambda_Z\,G_{Z_c\chi_{c0} \rho}}{\left(p^{\prime2}-m_Z^2\right)\left(p^2-m_{\chi_{c0}}^2 \right)\left(q^2-m_{\rho}^2\right)}\left(-g_{\mu\alpha}+\frac{q_{\mu}q_{\alpha}}{q^2}\right)\left(-g_\nu{}^\alpha+\frac{p^\prime_{\nu}p^{\prime\alpha}}{p^{\prime2}}\right)+\cdots\nonumber\\
&=&\Pi(p^{\prime2},p^2,q^2)\,g_{\mu\nu}+\cdots \, ,
\end{eqnarray}

\begin{eqnarray}
\Pi^5_{\alpha\beta\nu}(p,q)&=& \frac{f_{D^{*0}}m_{D^{*0}}f_{D^{*+}}m_{D^{*+}}\,\lambda_Z\,G_{Z_c\bar{D}^*D^*}}{\left(p^{\prime2}-m_Z^2\right)\left(p^2-m_{D^{*0}}^2 \right)\left(q^2-m_{D^{*+}}^2\right)}\left(p-q\right)^\sigma \left(-g_{\nu\sigma}+\frac{p^{\prime}_{\nu}p^\prime_{\sigma}}{p^{\prime2}}\right) \left(-g_{\alpha\rho} +\frac{p_{\alpha}p_{\rho}}{p^2}\right)\nonumber\\
&&\left(-g_{\beta}{}^\rho+\frac{q_{\beta}q^{\rho}}{q^2}\right)+\cdots\nonumber\\
&=&\Pi(p^{\prime2},p^2,q^2)\,\left( -g_{\alpha\beta}p_\nu\right)+\cdots \, ,
\end{eqnarray}

\begin{eqnarray}
\Pi^6_{\nu}(p,q)&=& \frac{f_{D^{0}}m_{D^{0}}^2f_{D^{+}}m_{D^{+}}^2}{m_c^2}\frac{ \lambda_Z\,G_{Z_c\bar{D}D}}{\left(p^{\prime2}-m_Z^2\right)\left(p^2-m_{D^0}^2 \right)\left(q^2-m_{D^+}^2\right)}\left(p-q\right)^\alpha\left(-g_{\alpha\nu}+\frac{p^{\prime}_{\alpha}p^\prime_{\nu}}{p^{\prime2}}\right)+\cdots\nonumber\\
&=&\Pi(p^{\prime2},p^2,q^2)\,\left( -p_{\nu}\right)+\cdots \, ,
\end{eqnarray}

\begin{eqnarray}
\Pi^7_{\mu\nu}(p,q)&=& \frac{f_{D^{+}}m_{D^{+}}^2}{m_c}\frac{f_{D^{*0}}m_{D^{*0}}\,\lambda_Z\,G_{Z_c\bar{D}^*D}\,\varepsilon_{\alpha\beta\rho\sigma}p^\alpha p^{\prime\rho}}{\left(p^{\prime2}-m_Z^2\right)\left(p^2-m_{D^{*0}}^2 \right)\left(q^2-m_{D^+}^2\right)}\left(-g_{\mu}{}^{\beta}+\frac{p_{\mu}p^{\beta}}{p^2}\right)\left(-g_{\nu}{}^{\sigma}+\frac{p^\prime_{\nu}p^{\prime\sigma}}{p^{\prime2}}\right)+\cdots\nonumber\\
&=&\Pi(p^{\prime2},p^2,q^2)\,\left(-\varepsilon_{\mu\nu\alpha\beta}p^{\alpha}q^{\beta}\right)+\cdots \, ,
\end{eqnarray}
where we have used the following definitions for the hadronic coupling constants,
\begin{eqnarray}
\langle J/\psi(p)\pi(q)|X(p^\prime)\rangle &=& i\,\varepsilon^{\alpha\beta\tau\sigma}\,p_\alpha \xi^{J/\psi*}_{\beta}p^{\prime}_{\tau}\xi_\sigma^{Z}\, G_{Z_cJ/\psi \pi}\, ,\nonumber\\
\langle \eta_c(p)\rho(q)|X(p^\prime)\rangle &=& i\,\varepsilon^{\alpha\beta\tau\sigma}\,q_\alpha \xi^{\rho*}_{\beta}p^{\prime}_\tau\xi_\sigma^{Z}\, G_{Z_c\eta_c \rho}\, ,\nonumber\\
\langle J/\psi(p)a_0(q)|X(p^\prime)\rangle &=& i\,\xi^{*\alpha}_{J/\psi}\xi_\alpha^{Z}\, G_{Z_cJ/\psi a_0}\, ,\nonumber\\
\langle \chi_{c0}(p)\rho(q)|X(p^\prime)\rangle &=& i\,\xi^{*\alpha}_{\rho}\xi_\alpha^{Z}\, G_{Z_c\chi_{c0} \rho}\, ,\nonumber\\
\langle \bar{D}^*(p)D^*(q)|X(p^\prime)\rangle &=& i\,(p-q)^\alpha\xi_\alpha^{Z}\xi^{\bar{D}^* *}_\beta\xi^{D^* *\beta}\, G_{Z_c\bar{D}^*D^*}\, ,\nonumber\\
\langle \bar{D}(p)D(q)|X(p^\prime)\rangle &=& i\,(p-q)^\alpha\xi_\alpha^{Z}\, G_{Z_c\bar{D}D}\, ,\nonumber\\
\langle \bar{D}^*(p)D(q)|X(p^\prime)\rangle &=& i\,\varepsilon^{\alpha\beta\tau\sigma}\,p_\alpha \xi^{\bar{D}^* *}_{\beta}p^{\prime}_\tau\xi_\sigma^{Z}\, G_{Z_c\bar{D}^*D }\, ,
\end{eqnarray}
where the $G_{Z_cJ/\psi \pi}$,   $ G_{Z_c\eta_c \rho}$, $G_{Z_cJ/\psi a_0}$,  $G_{Z_c\chi_{c0} \rho}$, $G_{Z_c\bar{D}^*D^*}$,  $G_{Z_c\bar{D}D}$ and $G_{Z_c\bar{D}^*D}$
   are the hadronic coupling constants.

We study the correlation functions $\Pi(p^{\prime2},p^2,q^2)$ at the QCD side, and  carry out the operator product expansion up to the vacuum condensates of dimension 5 by taking   into account both the connected and disconnected Feynman diagrams and neglect the tiny contributions of the gluon condensate. In Fig.1, we draw the Feynman diagrams for the correlation functions  $\Pi^{i}_{\mu\nu}(p,q)$ with $i=1,\,2,\,3,\,4$ as an example. The connected and disconnected Feynman diagrams correspond to the
 factorizable  and non-factorizable contributions respectively. The  factorizable contributions, even  the perturbative terms, involve the rearrangements  in the color, flavor, and Dirac spinor spaces, which differ  greatly from the fall-apart decay of a loosely  bound two-meson state.
\begin{figure}
 \centering
  \includegraphics[totalheight=3cm,width=4cm]{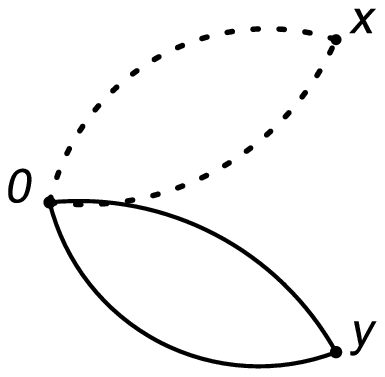}
 \includegraphics[totalheight=3cm,width=4cm]{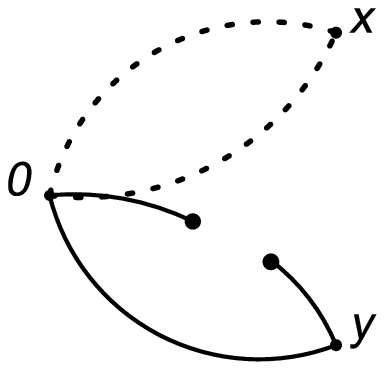}
\vglue+3mm
 \includegraphics[totalheight=3cm,width=4cm]{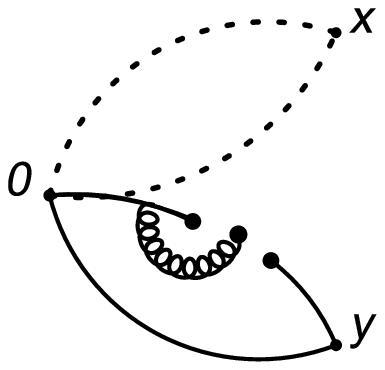}
 \includegraphics[totalheight=3cm,width=4cm]{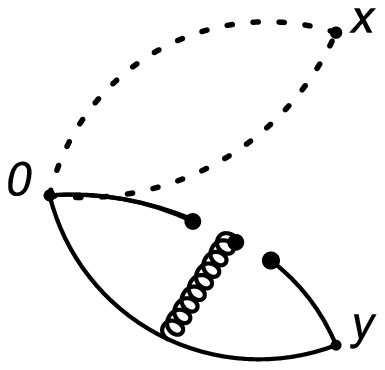}
 \vglue+3mm
 \includegraphics[totalheight=3cm,width=4cm]{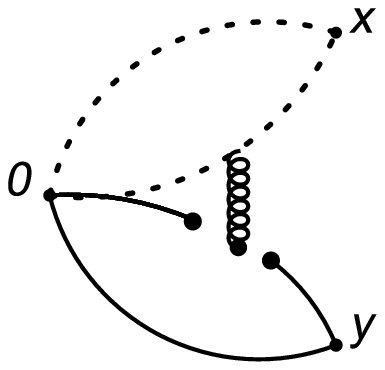}
 \caption{ The   Feynman diagrams  for the correlation functions $\Pi^{i}_{\mu\nu}(p,q)$ with $i=1,\,2,\,3,\,4$ , where the solid lines and dashed lines denote the light quarks and heavy quarks, respectively, the waved lines denote the gluons. Other diagrams obtained  by interchanging of the light or heavy quark lines are implied.  }
\end{figure}

Now we obtain the hadronic spectral densities and QCD spectral densities through dispersion relation, and write down the correlation functions in the spectral representation,
\begin{eqnarray}
\Pi_H(p^{\prime2},p^2,q^2)&=&\int_{\Delta^2}^{\infty} ds^\prime \int_{\Delta_s^2}^{\infty} ds \int_{\Delta_u^2}^{\infty} du \frac{\rho_{H}(s^{\prime},s,u)}{(s^\prime-p^{\prime2})(s-p^2)(u-q^2)}\, , \nonumber\\
\Pi_{QCD}(p^{\prime2},p^2,q^2)&=& \int_{\Delta_s^2}^{\infty} ds \int_{\Delta_u^2}^{\infty} du \frac{\rho_{QCD}(s,u)}{(s-p^2)(u-q^2)}\, ,
\end{eqnarray}
where the subscripts $H$ and $QCD$ denote the hadron side and QCD side of the correlation functions respectively,  $\Delta^2=(m_{B}+m_{C})^2$ (more precisely $\Delta^2=(\Delta_s+\Delta_u)^2$), the $\Delta_s^2$ and $\Delta_u^2$ are the thresholds. There are three  variables $s^\prime$, $s$ and $u$ at the hadron side, while there are two  variables  $s$ and $u$ at the QCD side.
We  math the hadron side  with the QCD side of the correlation functions,
and carry out the integral over $ds^\prime$  firstly to obtain the solid duality  for the decays $Z \to BC$ \cite{Wang-Y4660-Decay,WangZhang-Solid},
\begin{eqnarray}
\int_{\Delta_s^2}^{s_B^0} ds \int_{\Delta_u^2}^{u_C^0} du \frac{\rho_{QCD}(s,u)}{(s-p^2)(u-q^2)}&=&\int_{\Delta_s^2}^{s_B^0} ds \int_{\Delta_u^2}^{u_C^0} du \frac{1}{(s-p^2)(u-q^2)}\left[ \int_{\Delta^2}^{\infty} ds^\prime \frac{\rho_{H}(s^{\prime},s,u)}{s^\prime-p^{\prime2}}\right]\, , \nonumber\\
\end{eqnarray}
where the $B$ and $C$ denote the final states, the  $s_{B}^0$ and $u_{C}^0$ are the continuum thresholds. Compared to other works on the two-body strong decays of the tetraquark state candidates \cite{Nielsen-Decay-2006,Nielsen-Decay-2013,Azizi-Decay-2016-1,Azizi-Decay-2016-2}, we do not need to introduce the continuum threshold parameter $s^0_{Z}$ in the $s^\prime$ channel  by hand to avoid contamination (such as introducing  $\frac{1}{s_Z^0-p^{\prime2}}$, setting $s_B^0=s_Z^0$), or make special assumption of the value of the squared momentum $q^2$ (such as taking the limit $q^2\to 0$ to obtain the QCD sum rules, calculating the $G_{Z_cBC}(Q^2=-q^2)$ at large $Q^2$ then   extracting the $G_{Z_cBC}(Q^2)$ to the physical region $Q^2=-m_C^2$ with highly model dependent functions). In other words,  we need only carry out the operator product expansion at large space-like region $-p^2\to\infty$ and $-q^2\to\infty$, where the operator product expansion works. For the technical details, one can consult Refs.\cite{Wang-Y4660-Decay,WangZhang-Solid}.
We write down the integral over $ds^\prime$ explicitly,
\begin{eqnarray}
\int_{\Delta^2}^{\infty} ds^\prime \frac{\rho_{H}(s^{\prime},s,u)}{s^\prime-p^{\prime2}}&=&\frac{\rho_H(s,u)}{m_Z^2-p^{\prime2}}+\int_{s^0_{Z}}^\infty dt\frac{ \rho_{Z^\prime B}(t,p^2,q^2)+\rho_{Z^\prime C}(t,p^2,q^2)}{t-p^{\prime2}}\, ,
\end{eqnarray}
where the ground state hadronic spectral densities $\rho_H(s,u)$ are known, the transitions $\rho_{Z^\prime B}(t,p^2,q^2)+\rho_{Z^\prime C}(t,p^2,q^2)$ between the higher resonances (or continuum states) and the ground states $B$, $C$ are unknown, we have to introduce the parameters $C_{Z^\prime B}$ and $C_{Z^\prime C}$ to parameterize the net effects,
\begin{eqnarray}
C_{Z^\prime B}&=&\int_{s^0_{Z}}^\infty dt\frac{ \rho_{Z^\prime B}(t,p^2,q^2)}{t-p^{\prime2}}\, ,\nonumber\\
C_{Z^\prime C}&=&\int_{s^0_{Z}}^\infty dt\frac{ \rho_{Z^\prime C}(t,p^2,q^2)}{t-p^{\prime2}}\, .
\end{eqnarray}
Then we set $p^{\prime^2}=p^2$, or $4p^2$, and perform the double Borel transform  with respect to  $P^2=-p^2$ and $Q^2=-q^2$ respectively  to obtain the  QCD sum rules for the hadronic coupling constants, which are given explicitly in the Appendix.

 Now  we give an example to illustrate how to carry out the operator product expansion. The correlation function $\Pi^1_{\mu\nu}(p)$ at the QCD side can be written as,
\begin{eqnarray}
\Pi_{\mu\nu}^1(p,q)&=&-\frac{i}{\sqrt{2}} \varepsilon^{ijk}\varepsilon^{imn}   \int d^4x d^4y e^{ip \cdot x} e^{iq \cdot y}   \nonumber\\
&&\Big\{{\rm Tr}\Big[ \gamma_{\mu} S_c^{ak}(x)  CS^{T}_{bj}(y)C \gamma_5 CS^{T}_{mb}(-y)C\gamma_\nu S_c^{na}(-x)\Big]   \nonumber\\
 &&+{\rm Tr}\Big[ \gamma_{\mu} S_c^{ak}(x) \gamma_\nu CS^{T}_{bj}(y)C \gamma_5 CS^{T}_{mb}(-y)C  S_c^{na}(-x)\Big]\Big\} \, ,
\end{eqnarray}
where the $i$, $j$, $k$, $m$, $\cdots$ are color indexes, the $S_c^{ak}(x)$ and $S^{bj}(y)$ are the full $c$ and $u/d$ quark propagators, respectively \cite{Reinders85,Pascual-1984,WangHuangTao-3900}.  We carry out the integrals over $d^4x$ and $d^4y$, and take into account all terms proportional to  the tensor structure $\varepsilon_{\mu\nu\alpha\beta}p^{\alpha}q^{\beta}$, irrespective of the perturbative terms, quark condensate terms and mixed quark condensates terms, in other words, we calculate all the Feynman diagrams shown in Fig.1.   However, not all terms (or Feynman diagrams) have contributions due to the special tensor structure $\varepsilon_{\mu\nu\alpha\beta}p^{\alpha}q^{\beta}$.

\section{Numerical results and discussions}
At the QCD side, we take the vacuum condensates  to be the standard values
$\langle\bar{q}q \rangle=-(0.24\pm 0.01\, \rm{GeV})^3$,   $\langle\bar{q}g_s\sigma G q \rangle=m_0^2\langle \bar{q}q \rangle$,
$m_0^2=(0.8 \pm 0.1)\,\rm{GeV}^2$   at the energy scale  $\mu=1\, \rm{GeV}$
\cite{SVZ79-1,SVZ79-2,Reinders85,ColangeloReview}, and  take the $\overline{MS}$ masses $m_{c}(m_c)=(1.275\pm0.025)\,\rm{GeV}$ and $m_s(\mu=2\,\rm{GeV})=0.095\,\rm{GeV}$
 from the Particle Data Group \cite{PDG}.
Moreover,  we take into account
the energy-scale dependence of  the quark condensate, mixed quark condensate and $\overline{MS}$ masses from the renormalization group equation,
 \begin{eqnarray}
 \langle\bar{q}q \rangle(\mu)&=&\langle\bar{q}q \rangle(Q)\left[\frac{\alpha_{q}(Q)}{\alpha_{q}(\mu)}\right]^{\frac{12}{33-2n_f}}\, , \nonumber\\
 \langle\bar{q}g_s \sigma G q \rangle(\mu)&=&\langle\bar{q}g_s \sigma G q \rangle(Q)\left[\frac{\alpha_{s}(Q)}{\alpha_{s}(\mu)}\right]^{\frac{2}{33-2n_f}}\, ,\nonumber\\
m_c(\mu)&=&m_c(m_c)\left[\frac{\alpha_{s}(\mu)}{\alpha_{s}(m_c)}\right]^{\frac{12}{33-2n_f}} \, ,\nonumber\\
m_s(\mu)&=&m_s({\rm 2GeV} )\left[\frac{\alpha_{s}(\mu)}{\alpha_{s}({\rm 2GeV})}\right]^{\frac{12}{33-2n_f}}\, ,\nonumber\\
\alpha_s(\mu)&=&\frac{1}{b_0t}\left[1-\frac{b_1}{b_0^2}\frac{\log t}{t} +\frac{b_1^2(\log^2{t}-\log{t}-1)+b_0b_2}{b_0^4t^2}\right]\, ,
\end{eqnarray}
  where $t=\log \frac{\mu^2}{\Lambda^2}$, $b_0=\frac{33-2n_f}{12\pi}$, $b_1=\frac{153-19n_f}{24\pi^2}$, $b_2=\frac{2857-\frac{5033}{9}n_f+\frac{325}{27}n_f^2}{128\pi^3}$,  $\Lambda=210\,\rm{MeV}$, $292\,\rm{MeV}$  and  $332\,\rm{MeV}$ for the flavors  $n_f=5$, $4$ and $3$, respectively \cite{PDG,Narison-mix}, and evolve all the input parameters to the optimal  energy scale   $\mu$  with $n_f=4$ to extract the hadronic coupling constants.
  In this article, we take the energy scales of the QCD spectral densities to be $\mu=\frac{m_{\eta_c}}{2}=1.5\,\rm{GeV}$, which is acceptable for the mesons $D$ and $J/\psi$ \cite{Wang-Y4660-Decay,WangHuangTao-3900}. In this article, we neglect the small $u$ and $q$ quark masses.

The hadronic parameters are chosen  as
$m_{J/\psi}=3.0969\,\rm{GeV}$, $m_{\pi}=0.13957\,\rm{GeV}$,  $m_{\rho}=0.77526\,\rm{GeV}$,
$m_{\eta_c}=2.9839\,\rm{GeV}$, $m_{a_0}=0.980\,\rm{GeV}$,
$m_{D^+}=1.8695\,\rm{GeV}$, $m_{D^0}=1.86484\,\rm{GeV}$, $m_{D^{*+}}=2.01026\,\rm{GeV}$, $m_{D^{*0}}=2.00685\,\rm{GeV}$,
$m_{\chi_{c0}}=3.41471\,\rm{GeV}$, $\sqrt{s^0_{D}}=2.4\,\rm{GeV}$, $\sqrt{s^0_{D^*}}=2.5\,\rm{GeV}$, $\sqrt{s^0_{J/\psi}}=3.6\,\rm{GeV}$, $\sqrt{s^0_{\eta_c}}=3.5\,\rm{GeV}$, $\sqrt{s^0_{\chi_{c0}}}=3.9\,\rm{GeV}$ \cite{PDG},
$f_{J/\psi}=0.418 \,\rm{GeV}$, $f_{\eta_c}=0.387 \,\rm{GeV}$  \cite{Becirevic},
$f_{\rho}=0.215\,\rm{GeV}$, $\sqrt{s^0_{\rho}}=1.3\,\rm{GeV}$   \cite{Wang-Zc4200},
$f_{a_0}=0.214\,\rm{GeV}$, $\sqrt{s^0_{a_0}}=1.3\,\rm{GeV}$   \cite{Wang-f980-decay},
$f_{D}=0.208\,\rm{GeV}$, $f_{D^*}=0.263\,\rm{GeV}$  \cite{Wang-heavy-decay},
$f_{\chi_{c0}}=0.359\,\rm{GeV}$ \cite{Charmonium-PRT},
$m_Z=4.59\,\rm{GeV}$ \cite{WangY4360Y4660-1803},   $\lambda_{Z}=6.21\times 10^{-2}\,\rm{GeV}^5$ \cite{WangY4360Y4660-1803},
   $f_{\pi}m^2_{\pi}/(m_u+m_d)=-2\langle \bar{q}q\rangle/f_{\pi}$ from the Gell-Mann-Oakes-Renner relation.

  We set the Borel parameters to be $T_1^2=T_2^2=T^2$ for simplicity.
The unknown parameters are chosen as
$C_{Z^{\prime}J/\psi}+C_{Z^{\prime}\pi}=-0.0014\,\rm{GeV}^6 $,
$C_{Z^{\prime}\eta_c}+C_{Z^{\prime}\rho}=0.00135\,\rm{GeV}^6 $,
$C_{Z^{\prime}J/\psi}+C_{Z^{\prime}a_0}=-0.0119 \,\rm{GeV}^8 $,
$C_{Z^{\prime}\chi_{c0}}+C_{Z^{\prime}\rho}=0.0066\,\rm{GeV}^8 $,
$C_{Z^{\prime}D^*}+C_{Z^{\prime}\bar{D}^*}=0.0031\,\rm{GeV}^7 $,
$C_{Z^{\prime}D}+C_{Z^{\prime}\bar{D}}=0.0022\,\rm{GeV}^7 $,
$C_{Z^{\prime}D}+C_{Z^{\prime}\bar{D}^*}=0.0003 \,\rm{GeV}^6 $
   to obtain  platforms in the Borel windows, which are shown in Table 1. The Borel windows $T_{max}^2-T^2_{min}=1.0\,\rm{GeV}^2$ for the charmonium decays
   and  $T_{max}^2-T^2_{min}=0.7\,\rm{GeV}^2$ for the open-charm decays, where the $T^2_{max}$ and $T^2_{min}$ denote the maximum and minimum of the Borel parameters.
We choose  the same intervals $T_{max}^2-T^2_{min}$ in all the QCD sum rules for the two-body strong decays \cite{Wang-Y4660-Decay,Wang-Y4140-Y4274-1,Wang-Y4140-Y4274-2}, which work well for the decays of the $X(4140)$, $X(4274)$ and $Y(4660)$.
  In Figs.2-3, we plot the hadronic coupling constants  $G$  with variations of the  Borel parameters $T^2$ at much larger intervals than the Borel windows. From the figures, we can see that there appear platforms in the Borel windows indeed.

\begin{table}
\begin{center}
\begin{tabular}{|c|c|c|c|c|c|c|c|}\hline\hline
                                      &$T^2(\rm{GeV}^2)$   &$|G|$                                    &$\Gamma(\rm{MeV})$   \\ \hline

$Z_c^-(4600)\to J/\psi \pi^-$         &$4.0-5.0$           &$0.90^{+0.20}_{-0.18}\,\rm{GeV}^{-1}$    &$41.4^{+20.5}_{-14.9}$     \\ \hline

$Z_c^-(4600)\to \eta_c\rho^-$         &$4.1-5.1$           &$1.01^{+0.34}_{-0.32}\,\rm{GeV}^{-1}$    &$41.6^{+32.7}_{-22.2}$     \\ \hline

$Z_c^-(4600)\to J/\psi a_0^-(980)$    &$3.4-4.4$           &$2.37^{+1.07}_{-0.97}\,\rm{GeV}$         &$10.2^{+11.3}_{-6.7}$     \\ \hline

$Z_c^-(4600)\to \chi_{c0}\rho^-$      &$3.1-4.1$           &$1.35^{+0.95}_{-0.82}\,\rm{GeV}$         &$3.5^{+6.7}_{-3.0}$     \\ \hline

$Z_c^-(4600)\to D^{*0} D^{*-}$        &$2.1-2.8$           &$1.58^{+0.51}_{-0.45} $                  &$39.5^{+29.6}_{-19.3}$     \\ \hline

$Z_c^-(4600)\to D^{0} D^{-}$          &$1.7-2.4$           &$1.05^{+0.32}_{-0.28} $                  &$6.6^{+4.6}_{-3.0}$     \\ \hline

$Z_c^-(4600)\to D^{*0} D^{-}$         &$1.7-2.4$           &$0.14^{+0.06}_{-0.06}\,\rm{GeV}^{-1}$    &$1.0^{+1.0}_{-0.6}$     \\ \hline\hline
\end{tabular}
\end{center}
\caption{ The Borel  windows, hadronic coupling constants,  partial decay widths of the $Z_c(4600)$. }
\end{table}

 We take into account the uncertainties of the input parameters, and obtain the  hadronic coupling constants, which are  shown in Table 1 and Figs.2-3.
 Now it is easy to obtain the partial decay widths of the two-body strong decays $Z_c(4600) \to J/\psi \pi$,   $\eta_c \rho$, $J/\psi a_0$,  $\chi_{c0} \rho$, $D^*\bar{D}^*$,  $D\bar{D}$, $D^*\bar{D}$ and $D\bar{D}^*$, which are also shown in Table 1. From Table 1, we can see that the partial decay width $\Gamma(Z_c^-(4600)\to J/\psi \pi^-)=41.4^{+20.5}_{-14.9}\,\rm{MeV}$ is rather large, which can account for the observation  of the $Z_c(4600)$ in the $J/\psi \pi^-$ mass spectrum,
 although  it is just an evidence.

\begin{figure}
 \centering
  \includegraphics[totalheight=5cm,width=7cm]{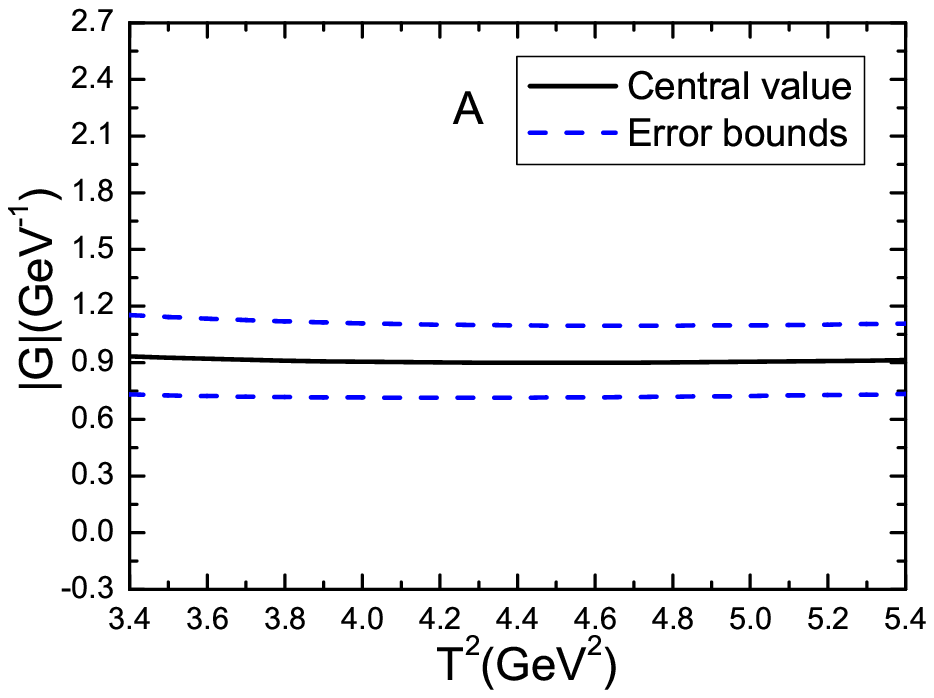}
  \includegraphics[totalheight=5cm,width=7cm]{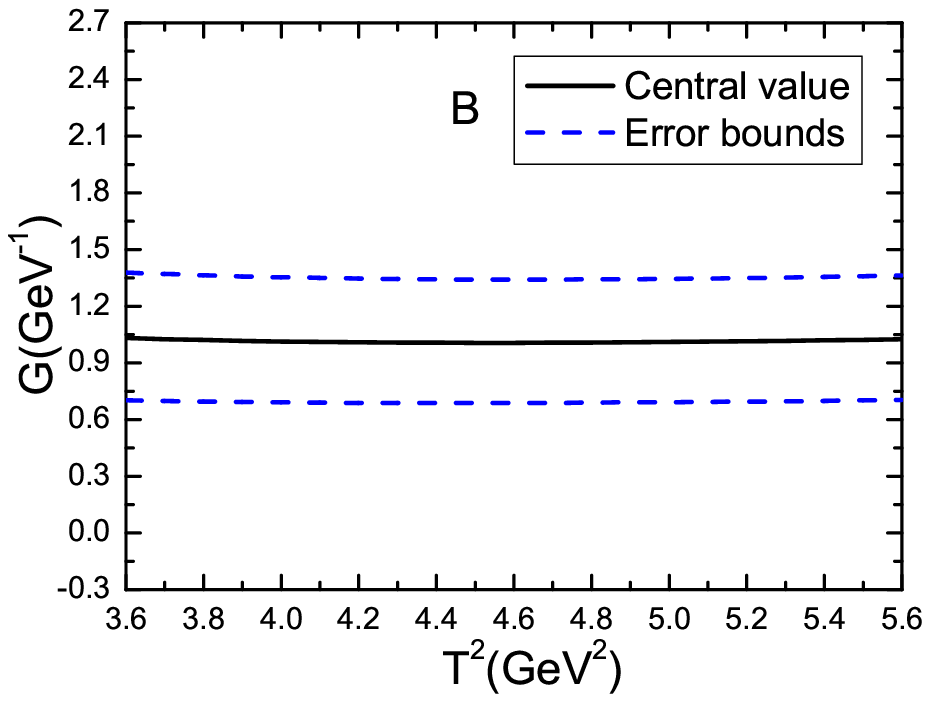}
   \includegraphics[totalheight=5cm,width=7cm]{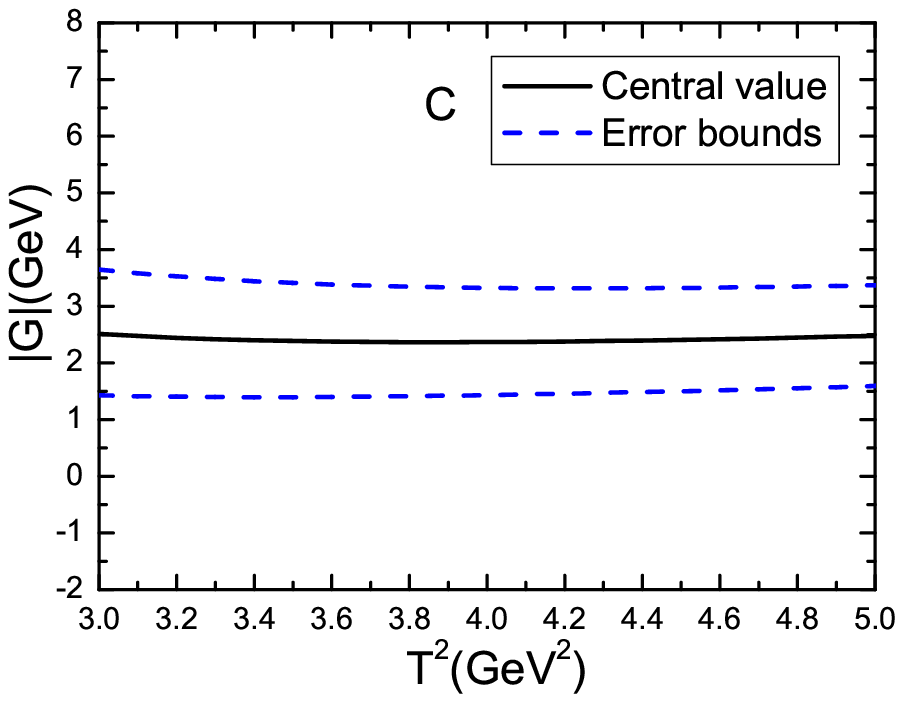}
  \includegraphics[totalheight=5cm,width=7cm]{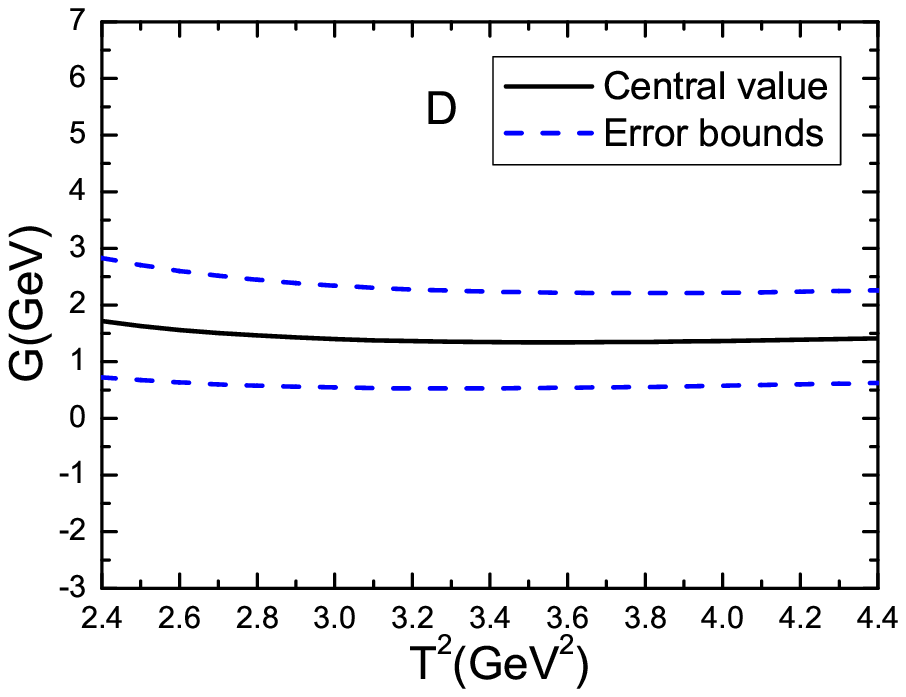}
     \caption{The hadronic coupling constants  with variations of the  Borel  parameters  $T^2$, where the $A$, $B$, $C$ and $D$  denote the
      $G_{Z_cJ/\psi \pi}$,   $ G_{Z_c\eta_c \rho}$, $G_{Z_cJ/\psi a_0}$ and  $G_{Z_c\chi_{c0} \rho}$,  respectively.}
\end{figure}

\begin{figure}
 \centering
  \includegraphics[totalheight=5cm,width=7cm]{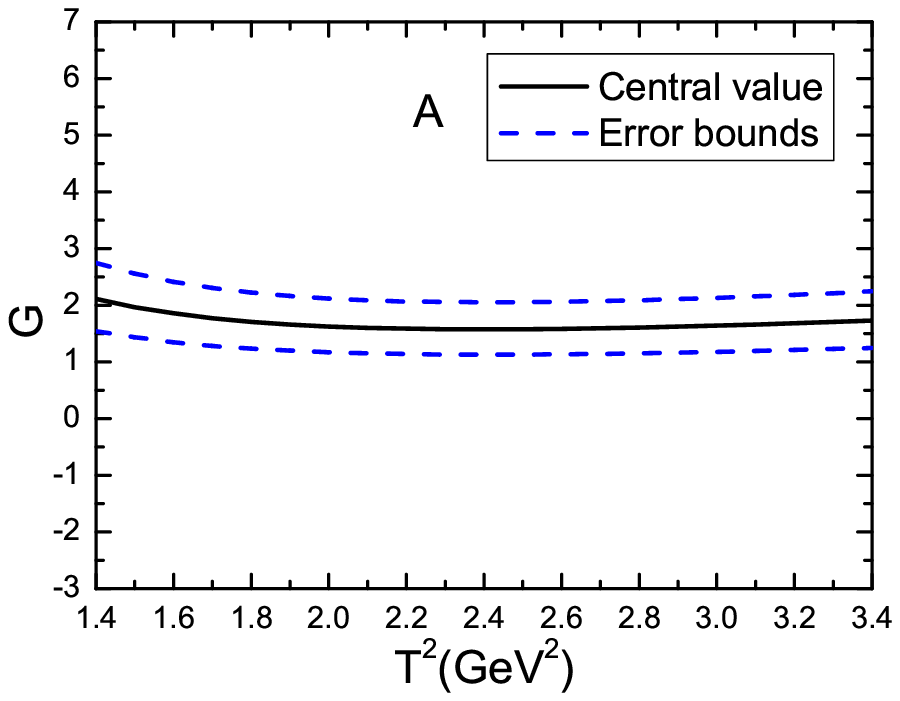}
  \includegraphics[totalheight=5cm,width=7cm]{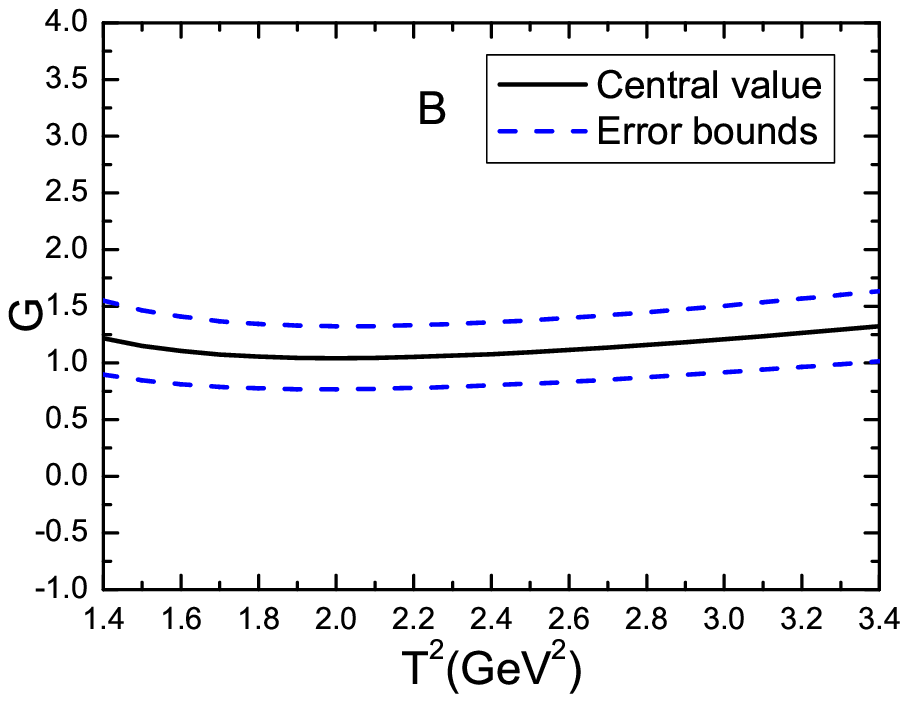}
   \includegraphics[totalheight=5cm,width=7cm]{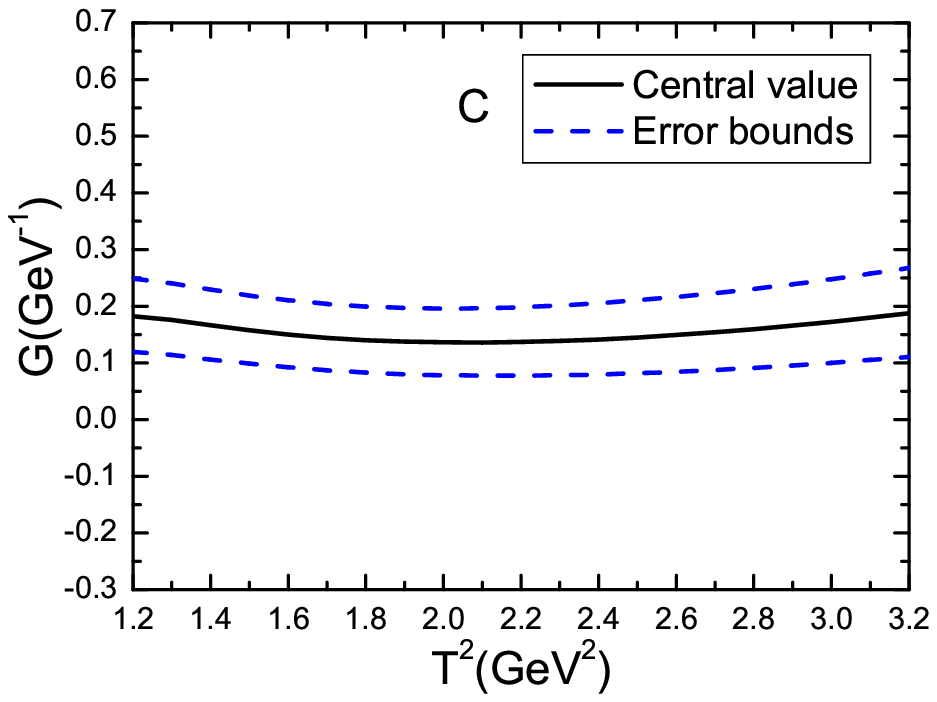}
     \caption{The hadronic coupling constants  with variations of the  Borel  parameters  $T^2$, where the $A$, $B$ and $C$  denote the
     $G_{Z_c\bar{D}^*D^*}$,  $G_{Z_c\bar{D}D}$ and $G_{Z_c\bar{D}^*D}$, respectively.}
\end{figure}

We can saturate the width with the two-body strong decays and obtain the total decay width,
\begin{eqnarray}
\Gamma(Z_c(4600) )&=& 144.8^{+50.6}_{-33.9}\,{\rm{MeV}}\, ,
\end{eqnarray}
which is reasonable for the tetraquark state. The present predictions can be confronted to the experimental data in the future, which may shed light on the nature of the $Z_c(4600)$.
The two possible  structures near $m(J/\psi \pi^-)=4200 \,\rm{MeV}$ and $4600\,\rm{MeV}$  have not been confirmed by other experiments, and their quantum numbers, such as the spin-parity $J^P$, have not  been measured yet. They maybe originate from statistical fluctuations, unsuitable cuts or subtractions, etc,   and do not exist at all. In previous work \cite{WangY4360Y4660-1803}, we observed  the $[qc]_P[\bar{q}\bar{c}]_A-[qc]_A[\bar{q}\bar{c}]_P$  type tetraquark state with the  $J^{PC}=1^{--}$ had a mass $(4.59 \pm 0.08)\,\rm{GeV}$, which happens to coincide   with the LHCb data. It is interesting  to assign the $Z_c(4600)$ to be $[qc]_P[\bar{q}\bar{c}]_A-[qc]_A[\bar{q}\bar{c}]_P$  type vector tetraquark state, and study its two-body strong decays to explore its structures. In this article, we obtain the partial decay widths of the  $[qc]_P[\bar{q}\bar{c}]_A-[qc]_A[\bar{q}\bar{c}]_P$  type   tetraquark state with $J^{PC}=1^{--}$, which are valuable in  both theoretical and experimental exploring the  structures and properties  of the tetraquark states, even if the $Z_c(4600)$ does not exist.

Now we make a crude estimation for the systematic uncertainties of the present QCD sum rules. For the correlation functions $\Pi^i_{\mu\nu}(p,q)$ with $i=1,\,2,\,3,\,4$, we set $p^{\prime2}=p^2$, while for the correlation functions $\Pi^i_{\alpha\beta\nu/\nu/\mu\nu}(p,q)$ with $i=4,\,5,\,6$, we set $p^{\prime2}=4p^2$, then perform the Borel transform with respect to $P^2=-p^2$ to obtain the QCD sum rules.
In fact, we can set $p^{\prime2}=\alpha^2 p^2$ by introducing an additional parameter $\alpha$, for example, the left side of the QCD sum rules in Eq.\eqref{QCDSR-pi} is changed to
 \begin{eqnarray}
&& \frac{f_{\pi}m_{\pi}^2}{m_u+m_d}\frac{f_{J/\psi}m_{J/\psi}\,\lambda_Z\,G_{Z_cJ/\psi\pi}\,}{\alpha^2\left(\widetilde{m}_Z^2-m_{J/\psi}^2\right)}\left[\exp\left(-\frac{m_{J/\psi}^2}{T_1^2}\right) -\exp\left(-\frac{\widetilde{m}_Z^2}{T_1^2}\right) \right]\exp\left( -\frac{m_{\pi}^2}{T_2^2}\right)\nonumber\\
 &&+\left(C_{Z^{\prime}J/\psi}+C_{Z^{\prime}\pi} \right)\exp\left(-\frac{m_{J/\psi}^2}{T_1^2} -\frac{m_{\pi}^2}{T_2^2}\right)\, ,
\end{eqnarray}
where $\widetilde{m}_Z^2=\frac{m_Z^2}{\alpha^2}$. If we choose $\alpha^2=1.2^2=1.44$, we can obtain the value $|G_{Z_cJ/\psi\pi}|=0.88\,\rm{GeV}^{-1}$ by choosing suitable $C_{Z^{\prime}J/\psi}+C_{Z^{\prime}\pi}$, the additional uncertainty is about $2\%$. $\frac{m_Z}{m_{J/\psi}}=1.48$, if we choose $\alpha^2>1.48$, say $\alpha^2=1.3^2=1.69$, no stable QCD sum rules can be obtained. Accordingly, if we take $\alpha=2.2$ ($\alpha^2=4.84$) to calculate the $G_{Z_cD\bar{D}}$, we can obtain a value $G_{Z_cD\bar{D}}=1.01$, the additional uncertainty is about $4\%$.  In the heavy quark limit, we can make a crude approximation $m_Z=m_{J/\psi}=2m_{D/D^*}$, the ideal values of the $\alpha$ for the correlation functions $\Pi^i_{\mu\nu/\alpha\beta\nu/\nu}(p,q)$ with $i=1,\,2,\,3,\,4$  and $5,\,6,\,7$ are $\alpha=1$ and $2$, respectively. In practical calculations, we observe that if there are larger  deviations  from $\alpha=1$ or $2$, no stable QCD sum rules can be obtained.

\section{Conclusion}
In this article, we tentatively assign the $Z_c(4600)$ to be the $[dc]_P[\bar{u}\bar{c}]_A-[dc]_A[\bar{u}\bar{c}]_P$ type vector tetraquark state, and obtain the QCD sum rules  for the hadronic coupling constants  $G_{Z_cJ/\psi \pi}$,   $ G_{Z_c\eta_c \rho}$, $G_{Z_cJ/\psi a_0}$,  $G_{Z_c\chi_{c0} \rho}$, $G_{Z_c\bar{D}^*D^*}$,  $G_{Z_c\bar{D}D}$ and $G_{Z_c\bar{D}^*D}$   based on solid quark-hadron duality by taking   into account both the connected and disconnected Feynman diagrams in the operator product expansion. The predictions for the partial decay widths
$\Gamma(Z_c^-\to J/\psi \pi^-)=41.4^{+20.5}_{-14.9}\,\rm{MeV}$,
$\Gamma(Z_c^-\to \eta_c\rho^-)=41.6^{+32.7}_{-22.2}\,\rm{MeV}$,
$\Gamma(Z_c^-\to J/\psi a_0^-(980))=10.2^{+11.3}_{-6.7}\,\rm{MeV}$,
$\Gamma(Z_c^-\to \chi_{c0}\rho^-)=3.5^{+6.7}_{-3.0}\,\rm{MeV}$,
$\Gamma(Z_c^-\to D^{*0} D^{*-})=39.5^{+29.6}_{-19.3}\,\rm{MeV}$,
$\Gamma(Z_c^-\to D^{0} D^{-})=6.6^{+4.6}_{-3.0}\,\rm{MeV}$ and
$\Gamma(Z_c^-\to D^{*0} D^{-})=1.0^{+1.0}_{-0.6}\,\rm{MeV}$
 can be confronted to the experimental data in the future to diagnose the nature of the  $Z_c(4600)$, as the LHCb collaboration have observed an evidence for  the $Z_c(4600)$ in the $J/\psi \pi^-$ mass spectrum.

\section*{Appendix}
The explicit expressions of the QCD sum rules for the hadronic coupling constants,
 \begin{eqnarray}\label{QCDSR-pi}
&& \frac{f_{\pi}m_{\pi}^2}{m_u+m_d}\frac{f_{J/\psi}m_{J/\psi}\,\lambda_Z\,G_{Z_cJ/\psi\pi}\,}{m_Z^2-m_{J/\psi}^2}\left[\exp\left(-\frac{m_{J/\psi}^2}{T_1^2}\right) -\exp\left(-\frac{m_Z^2}{T_1^2}\right) \right]\exp\left( -\frac{m_{\pi}^2}{T_2^2}\right)\nonumber\\
 &&+\left(C_{Z^{\prime}J/\psi}+C_{Z^{\prime}\pi} \right)\exp\left(-\frac{m_{J/\psi}^2}{T_1^2} -\frac{m_{\pi}^2}{T_2^2}\right)\nonumber\\
&=&\frac{m_c\langle\bar{q}q\rangle}{2\sqrt{2}\pi^2}\int_{4m_c^2}^{s^0_{J/\psi}} ds
\sqrt{1-\frac{4m_c^2}{s}}\exp\left(-\frac{s}{T_1^2}\right)\nonumber\\
&&+\frac{m_c\langle\bar{q}g_{s}\sigma Gq\rangle}{24\sqrt{2}\pi^2}\int_{4m_c^2}^{s^0_{J/\psi}} ds
\frac{1}{\sqrt{s\left(s-4m_c^2\right)}}\frac{s-2m_c^2}{s}\exp\left(-\frac{s}{T_1^2}\right) \, ,
\end{eqnarray}

\begin{eqnarray}
&& \frac{f_{\eta_c}m_{\eta_c}^2}{2m_c}\frac{f_{\rho}m_{\rho}\,\lambda_Z\,G_{Z_c\eta_c\rho}\,}{m_Z^2-m_{\eta_c}^2}\left[\exp\left(-\frac{m_{\eta_c}^2}{T_1^2}\right) -\exp\left(-\frac{m_Z^2}{T_1^2}\right) \right]\exp\left( -\frac{m_{\rho}^2}{T_2^2}\right)\nonumber\\
 &&+\left(C_{Z^{\prime}\eta_c}+C_{Z^{\prime}\rho} \right)\exp\left(-\frac{m_{\eta_c}^2}{T_1^2} -\frac{m_{\rho}^2}{T_2^2}\right)\nonumber\\
&=&-\frac{m_c\langle\bar{q}q\rangle}{2\sqrt{2}\pi^2}\int_{4m_c^2}^{s^0_{\eta_{c}}} ds
\sqrt{1-\frac{4m_c^2}{s}}\exp\left(-\frac{s}{T_1^2}\right)+\frac{m_c\langle\bar{q}g_{s}\sigma Gq\rangle}{6\sqrt{2}\pi^2T_2^2}\int_{4m_c^2}^{s^0_{\eta_{c}}} ds
\sqrt{1-\frac{4m_c^2}{s}}\exp\left(-\frac{s}{T_1^2}\right)\nonumber\\
&&-\frac{m_c\langle\bar{q}g_{s}\sigma Gq\rangle}{24\sqrt{2}\pi^2}\int_{4m_c^2}^{s^0_{\eta_{c}}} ds
\frac{1}{\sqrt{s\left(s-4m_c^2\right)}}\exp\left(-\frac{s}{T_1^2}\right) \, ,
\end{eqnarray}

\begin{eqnarray}
 &&\frac{f_{J/\psi}m_{J/\psi}f_{a_0}m_{a_0}\,\lambda_Z\,G_{Z_cJ/\psi a_0}}{m_Z^2-m_{J/\psi}^2 } \left[\exp\left(-\frac{m_{J/\psi}^2}{T_1^2}\right) -\exp\left(-\frac{m_Z^2}{T_1^2}\right) \right]\exp\left( -\frac{m_{a_0}^2}{T_2^2}\right)\nonumber\\
 &&+\left(C_{Z^{\prime}J/\psi}+C_{Z^{\prime}a_0} \right)\exp\left(-\frac{m_{J/\psi}^2}{T_1^2} -\frac{m_{a_0}^2}{T_2^2}\right)\nonumber\\
&=&-\frac{1}{32\sqrt{2}\pi^4}\int_{4m_c^2}^{s^0_{J/\psi}} ds \int_0^{s^0_{a_0}} du us
\sqrt{1-\frac{4m_c^2}{s}}\left(1+\frac{2m_c^2}{s}\right)\exp\left(-\frac{s}{T_1^2}-\frac{u}{T_2^2}\right)\, ,
\end{eqnarray}

\begin{eqnarray}
&& \frac{f_{\chi_{c0}}m_{\chi_{c0}}f_{\rho}m_{\rho}\,\lambda_Z\,G_{Z_c\chi_{c0} \rho}}{m_Z^2-m_{\chi_{c0}}^2}\left[\exp\left(-\frac{m_{\chi_{c0}}^2}{T_1^2}\right) -\exp\left(-\frac{m_Z^2}{T_1^2}\right) \right]\exp\left( -\frac{m_{\rho}^2}{T_2^2}\right)\nonumber\\
 &&+\left(C_{Z^{\prime}\chi_{c0}}+C_{Z^{\prime}\rho} \right)\exp\left(-\frac{m_{\chi_{c0}}^2}{T_1^2} -\frac{m_{\rho}^2}{T_2^2}\right)\nonumber\\
&=&\frac{1}{32\sqrt{2}\pi^4}\int_{4m_c^2}^{s^0_{\chi_{c0}}} ds \int_0^{s^0_\rho} du u s
\sqrt{1-\frac{4m_c^2}{s}}\left(1-\frac{4m_c^2}{s}\right)\exp\left(-\frac{s}{T_1^2}-\frac{u}{T_2^2}\right)  \, ,
\end{eqnarray}

\begin{eqnarray}
&& \frac{f_{D^{*0}}m_{D^{*0}}f_{D^{*+}}m_{D^{*+}}\,\lambda_Z\,G_{Z_c\bar{D}^*D^*}}{4\left(\widetilde{m}_Z^2-m_{D^{*0}}^2\right)}
\left[\exp\left(-\frac{m_{D^{*0}}^2}{T_1^2}\right) -\exp\left(-\frac{\widetilde{m}_Z^2}{T_1^2}\right) \right]\exp\left( -\frac{m_{D^{*+}}^2}{T_2^2}\right)\nonumber\\
 &&+\left(C_{Z^{\prime}\bar{D}^*}+C_{Z^{\prime}D^*} \right)\exp\left(-\frac{m_{D^{*0}}^2}{T_1^2} -\frac{m_{D^{*+}}^2}{T_2^2}\right)\nonumber\\
&=&\frac{m_c}{64\sqrt{2}\pi^4}\int_{m_c^2}^{s^0_{ D^*}} ds \int_{m_c^2}^{s^0_{D^*}} du
\left(2u+m_c^2\right)\left(1-\frac{m_c^2}{s}\right)^2\left(1-\frac{m_c^2}{u}\right)^2\exp\left(-\frac{s}{T_1^2}-\frac{u}{T_2^2}\right) \nonumber\\
&&-\frac{\langle\bar{q}q\rangle}{24\sqrt{2}\pi^2}\int_{m_c^2}^{s^0_{D^*}} du \left(2u+m_c^2\right)\left(1-\frac{m_c^2}{u}\right)^2 \exp\left(-\frac{m_c^2}{T_1^2}-\frac{u}{T_2^2}\right) \nonumber\\
&&-\frac{m_c^2\langle\bar{q}q\rangle}{8\sqrt{2}\pi^2} \int_{m_c^2}^{s^0_{D^*}} ds \left(1-\frac{m_c^2}{s}\right)^2 \exp\left(-\frac{s}{T_1^2}-\frac{m_c^2}{T_2^2}\right) \nonumber\\
&&+\frac{\langle\bar{q}g_{s}\sigma Gq\rangle}{288\sqrt{2}\pi^2T_1^2}\int_{m_c^2}^{s^0_{D^*}} du
\left(2u+m_c^2\right)\left(1-\frac{m_c^2}{u}\right)^2\left(4+\frac{3m_c^2}{T_1^2}\right)\exp\left(-\frac{m_c^2}{T_1^2}-\frac{u}{T_2^2}\right) \nonumber\\
&&+\frac{m_c^4\langle\bar{q}g_{s}\sigma Gq\rangle}{32\sqrt{2}\pi^2T_2^4} \int_{m_c^2}^{s^0_{D^*}} ds \left(1-\frac{m_c^2}{s}\right)^2 \exp\left(-\frac{s}{T_1^2}-\frac{m_c^2}{T_2^2}\right) \nonumber\\
&&+\frac{\langle\bar{q}g_{s}\sigma Gq\rangle}{96\sqrt{2}\pi^2}\int_{m_c^2}^{s^0_{D^*}} du \frac{m_c^2}{u} \exp\left(-\frac{m_c^2}{T_1^2}-\frac{u}{T_2^2}\right) \nonumber\\
&&-\frac{\langle\bar{q}g_{s}\sigma Gq\rangle}{96\sqrt{2}\pi^2} \int_{m_c^2}^{s^0_{D^*}} ds \frac{m_c^2}{s}\left(1-\frac{m_c^2}{s}\right)\left(6-\frac{m_c^2}{s}\right) \exp\left(-\frac{s}{T_1^2}-\frac{m_c^2}{T_2^2}\right) \nonumber\\
&&+\frac{\langle\bar{q}g_{s}\sigma Gq\rangle}{96\sqrt{2}\pi^2} \int_{m_c^2}^{s^0_{D^*}} ds
\frac{m_c^4}{s^2}\exp\left(-\frac{s}{T_1^2}-\frac{m_c^2}{T_2^2}\right)\, ,
\end{eqnarray}

\begin{eqnarray}
&& \frac{f_{D^{0}}m_{D^{0}}^2f_{D^{+}}m_{D^{+}}^2}{m_c^2}\frac{ \lambda_Z\,G_{Z_c\bar{D}D}}{4\left(\widetilde{m}_Z^2-m_{D^0}^2\right)}\left[\exp\left(-\frac{m_{D^0}^2}{T_1^2}\right) -\exp\left(-\frac{\widetilde{m}_Z^2}{T_1^2}\right) \right]\exp\left( -\frac{m_{D^+}^2}{T_2^2}\right)\nonumber\\
 &&+\left(C_{Z^{\prime}\bar{D}}+C_{Z^{\prime}D} \right)\exp\left(-\frac{m_{D^0}^2}{T_1^2} -\frac{m_{D^+}^2}{T_2^2}\right)\nonumber\\
&=&\frac{3m_c}{64\sqrt{2}\pi^4}\int_{m_c^2}^{s^0_{D}} ds \int_{m_c^2}^{s^0_{D}} du u
\left(1-\frac{m_c^2}{s}\right)^2\left(1-\frac{m_c^2}{u}\right)^2\exp\left(-\frac{s}{T_1^2}-\frac{u}{T_2^2}\right) \nonumber\\
&&-\frac{\langle\bar{q}q\rangle}{8\sqrt{2}\pi^2}\int_{m_c^2}^{s^0_{D}} du u\left(1-\frac{m_c^2}{u}\right)^2\exp\left(-\frac{m_c^2}{T_1^2}-\frac{u}{T_2^2}\right) \nonumber\\
&&-\frac{m_c^2\langle\bar{q}q\rangle}{8\sqrt{2}\pi^2}\int_{m_c^2}^{s^0_{D}} ds\left(1-\frac{m_c^2}{s}\right)^2\exp\left(-\frac{s}{T_1^2}-\frac{m_c^2}{T_2^2}\right) \nonumber\\
&&+\frac{m_c^2\langle\bar{q}g_{s}\sigma Gq\rangle}{32\sqrt{2}\pi^2T_1^4}\int_{m_c^2}^{s^0_{D}} du u\left(1-\frac{m_c^2}{u}\right)^2\exp\left(-\frac{m_c^2}{T_1^2}-\frac{u}{T_2^2}\right)   \nonumber\\
&&-\frac{m_c^2\langle\bar{q}g_{s}\sigma Gq\rangle}{16\sqrt{2}\pi^2T_2^2}\int_{m_c^2}^{s^0_{D}} ds \left(1-\frac{m_c^2}{s}\right)^2 \left(1-\frac{m_c^2}{2T_2^2}\right)
\exp\left(-\frac{s}{T_1^2}-\frac{m_c^2}{T_2^2}\right) \nonumber\\
&&-\frac{\langle\bar{q}g_{s}\sigma Gq\rangle}{192\sqrt{2}\pi^2}\int_{m_c^2}^{s^0_{D}} du
\left(1-\frac{m_c^2}{u}\right)\left(3-\frac{m_c^2}{u}\right)\exp\left(-\frac{m_c^2}{T_1^2}-\frac{u}{T_2^2}\right) \nonumber\\
&&-\frac{\langle\bar{q}g_{s}\sigma Gq\rangle}{96\sqrt{2}\pi^2}\int_{m_c^2}^{s^0_{D}} ds
\frac{m_c^2}{s}\left(1-\frac{m_c^2}{s}\right)\left(6-\frac{m_c^2}{s}\right)\exp\left(-\frac{s}{T_1^2}-\frac{m_c^2}{T_2^2}\right) \nonumber\\
&&-\frac{\langle\bar{q}g_{s}\sigma Gq\rangle}{192\sqrt{2}\pi^2}\int_{m_c^2}^{s^0_{D}} du
\left(3-\frac{m_c^4}{u^2}\right)\exp\left(-\frac{m_c^2}{T_1^2}-\frac{u}{T_2^2}\right)\nonumber\\
&&-\frac{\langle\bar{q}g_{s}\sigma Gq\rangle}{96\sqrt{2}\pi^2}\int_{m_c^2}^{s^0_{D}} ds\frac{m_c^6}{s^3}\exp\left(-\frac{s}{T_1^2}-\frac{m_c^2}{T_2^2}\right) \, ,
\end{eqnarray}

\begin{eqnarray}
&& \frac{f_{D^{+}}m_{D^{+}}^2}{m_c}\frac{f_{D^{*0}}m_{D^{*0}}\,\lambda_Z\,G_{Z_c\bar{D}^*D} }{4\left(\widetilde{m}_Z^2-m_{D^{*0}}^2\right)}\left[\exp\left(-\frac{m_{D^{*0}}^2}{T_1^2}\right) -\exp\left(-\frac{\widetilde{m}_Z^2}{T_1^2}\right) \right]\exp\left( -\frac{m_{D^+}^2}{T_2^2}\right)\nonumber\\
 &&+\left(C_{Z^{\prime}\bar{D}^*}+C_{Z^{\prime}D} \right)\exp\left(-\frac{m_{D^{*0}}^2}{T_1^2} -\frac{m_{D^+}^2}{T_2^2}\right)\nonumber\\
&=&-\frac{\langle\bar{q}g_{s}\sigma Gq\rangle}{32\sqrt{2}\pi^2}\int_{m_c^2}^{s^0_{D}} du
\frac{m_c}{u}\left(1-\frac{m_c^2}{u}\right)\exp\left(-\frac{m_c^2}{T_1^2}-\frac{u}{T_2^2}\right) \nonumber\\
&&+\frac{\langle\bar{q}g_{s}\sigma Gq\rangle}{32\sqrt{2}\pi^2}\int_{m_c^2}^{s^0_{D^*}} ds
\frac{m_c}{s}\left(1-\frac{m_c^2}{s}\right)\exp\left(-\frac{s}{T_1^2}-\frac{m_c^2}{T_2^2}\right) \nonumber\\
&&-\frac{\langle\bar{q}g_{s}\sigma Gq\rangle}{48\sqrt{2}\pi^2}\int_{m_c^2}^{s^0_{D}} du
\frac{m_c^3}{u^2}\exp\left(-\frac{m_c^2}{T_1^2}-\frac{u}{T_2^2}\right) \, ,
\end{eqnarray}
where $\widetilde{m}_Z^2=\frac{m_Z^2}{4}$, the $T_1^2$ and $T_2^2$ are the Borel parameters,  the unknown functions $C_{Z^{\prime}J/\psi}+C_{Z^{\prime}\pi}$,
$C_{Z^{\prime}\eta_c}+C_{Z^{\prime}\rho}$,
$C_{Z^{\prime}J/\psi}+C_{Z^{\prime}a_0}$,
$C_{Z^{\prime}\chi_{c0}}+C_{Z^{\prime}\rho}$,
$C_{Z^{\prime}D^*}+C_{Z^{\prime}\bar{D}^*}$,
$C_{Z^{\prime}D}+C_{Z^{\prime}\bar{D}}$ and
$C_{Z^{\prime}D}+C_{Z^{\prime}\bar{D}^*}$ parameterize the higher resonances or continuum state contributions.
In calculations, we observe that there appears divergence due to the endpoint $s=4m_c^2$, we can avoid the endpoint divergence with the simple replacement  $\frac{1}{\sqrt{s-4m_c^2}} \to \frac{1}{\sqrt{s-4m_c^2+4m_s^2}}$  by adding a small squared $s$-quark mass $4m_s^2$ \cite{Wang-Y4660-Decay,Wang-Y4140-Y4274-1,Wang-Y4140-Y4274-2}.

\section*{Acknowledgements}
This  work is supported by National Natural Science Foundation, Grant Number  11775079.


\begin{thebibliography}{99}

\bibitem{LHCb-Z4600}  R. Aaij et al, Phys. Rev. Lett. {\bf 122} (2019) 152002.

\bibitem{Belle-4200} K. Chilikin et al, Phys. Rev. {\bf D90} (2014) 112009.

\bibitem{WangY4360Y4660-1803} Z. G. Wang, Eur. Phys. J. {\bf C78} (2018)  518.

\bibitem{Wang-tetra-formula}  Z. G. Wang, Eur. Phys. J. {\bf C74} (2014)  2874.

\bibitem{WangEPJC-1601} Z. G. Wang, Eur. Phys. J. {\bf C76} (2016)  387.

\bibitem{Wang-Y4660-Decay}  Z. G. Wang, Eur. Phys. J. {\bf C79} (2019)  184.

\bibitem{Belle-Y4660-2014} X. L. Wang et al, Phys. Rev. {\bf D91}   (2015) 112007.

\bibitem{WangZhang-Solid} Z. G. Wang and  J. X. Zhang, Eur. Phys. J. {\bf C78} (2018) 14.


\bibitem{WangScalarNonet}  Z. G. Wang,   Eur. Phys. J. {\bf C76} (2016) 427.

\bibitem{SVZ79-1} M. A. Shifman, A. I. Vainshtein and V. I. Zakharov, Nucl. Phys. {\bf B147} (1979) 385.

\bibitem{SVZ79-2} M. A. Shifman, A. I. Vainshtein and V. I. Zakharov, Nucl. Phys. {\bf B147} (1979) 448.

\bibitem{Reinders85} L. J. Reinders, H. Rubinstein and S. Yazaki, Phys. Rept. {\bf 127} (1985) 1.

\bibitem{Nielsen-Decay-2006} F. S. Navarra and M. Nielsen, Phys. Lett. {\bf B639} (2006) 272.

\bibitem{Nielsen-Decay-2013} J. M. Dias, F. S. Navarra, M. Nielsen and C. M. Zanetti, Phys. Rev. {\bf D88} (2013)  016004.

\bibitem{Azizi-Decay-2016-1} S. S. Agaev, K. Azizi and H. Sundu, Phys. Rev. {\bf D93} (2016) 074002.

\bibitem{Azizi-Decay-2016-2} S. S. Agaev, K. Azizi and H. Sundu, Phys. Rev. {\bf D96} (2017)  034026.



\bibitem{Pascual-1984} P. Pascual and R. Tarrach, ``QCD: Renormalization for the practitioner", Springer Berlin Heidelberg (1984).

\bibitem{WangHuangTao-3900} Z. G. Wang and T. Huang,  Phys. Rev. {\bf D89} (2014) 054019.

\bibitem{ColangeloReview} P. Colangelo and A. Khodjamirian, hep-ph/0010175.


\bibitem{PDG}  M. Tanabashi et al, Phys. Rev. {\bf  D98} (2018)  030001.


\bibitem{Narison-mix} S. Narison and R. Tarrach, Phys. Lett. {\bf 125 B} (1983) 217.



\bibitem{Becirevic} D. Becirevic, G. Duplancic, B. Klajn, B. Melic and F. Sanfilippo,  Nucl. Phys. {\bf B883} (2014) 306.

\bibitem{Wang-Zc4200}  Z. G. Wang, Int. J. Mod. Phys. {\bf A30} (2015)  1550168.

\bibitem{Wang-f980-decay}  Z. G. Wang, W. M. Yang and S. L. Wan, Eur. Phys. J. {\bf C37} (2004)  223.

\bibitem{Wang-heavy-decay} Z. G. Wang, Eur. Phys. J. {\bf C75} (2015) 427.

\bibitem{Charmonium-PRT} V. A. Novikov, L. B. Okun, M. A. Shifman, A. I. Vainshtein, M. B. Voloshin and V. I. Zakharov, Phys. Rept. {\bf 41} (1978) 1.

\bibitem{Wang-Y4140-Y4274-1}  Z. G. Wang and Z. Y. Di, Eur. Phys. J. {\bf C79} (2019)  72.

\bibitem{Wang-Y4140-Y4274-2} Z. G. Wang, arXiv:1812.04503.

\end{thebibliography}
\end{document}